\documentstyle[preprint,amssymb,prl,aps,epsf]{revtex}

\large

\begin{document}

\epsfverbosetrue
\input epsf.sty 
\title{Muffin Tin Orbitals of Arbitrary Order}
\author{O.K. Andersen and T. Saha-Dasgupta}
\address{Max-Planck-Institut f\"{u}r Festk\"{o}rperforschung,
Postfach 800665, D-70506 Stuttgart, Germany }
\date{\today}
\maketitle

\begin{abstract}
We have derived orbital basis sets from scattering theory. They are
expressed as polynomial approximations to the energy dependence of a
set of partial waves, in quantized form. The corresponding matrices, as well as 
the Hamiltonian and overlap matrices, are specified by
the values on the energy mesh of the screened resolvent
and its first energy derivative. These orbitals are a generalization
of the 3rd-generation linear MTOs and should be useful  for electronic-structure
calculations in general. 
\end{abstract}


For electrons in condensed matter, it is often desirable to express the
one-electron wave functions, $\Psi _{i}\left( {\bf r}\right) ,$ with
energies, $\varepsilon _{i},$ in a certain range in terms of a {\em minimal
set} of energy-independent orbitals, $\chi _{RL}\left( {\bf r}\right) .$
Here, $R$ labels sites and $L$ the local symmetry (e.g. $L{\rm \equiv }lm).$

The simplest example of such an orbital is the Wannier function, $\chi
\left( {\bf r-R}\right) ,$ for an isolated band. A more realistic example is
illustrated in Fig.\thinspace \ref{Figorb}, the conduction-band orbital of a
cuprate high-temperature superconductor. This orbital is centered on Cu, has
anti-bonding O$_{x}\,p_{x}\,$--\thinspace Cu$\,d_{x^{2}-y^{2}}\,$%
--\thinspace O$_{y}\,p_{y}$ character, and extends beyond the 3rd-nearest
neighbor atoms. Its Bloch sum describes a tight-binding (TB) band: $%
\varepsilon _{{\bf k}}\sim \left\langle \varepsilon \right\rangle -2t\left(
\cos k_{x}+\cos k_{y}\right) +4t^{\prime }\cos k_{x}\cos k_{y}-2t^{\prime
\prime }\left( \cos 2k_{x}+\cos 2k_{y}\right) .$ This orbital is the
starting point for descriptions of the low-energy physics of the cuprates.
Its is {\em not} a Wannier function. First of all because the conduction
band is merely one partner of a bonding, non-bonding, anti-bonding triple
with nearly degenerate Cu$\,d$ and O\thinspace $p$ levels so that the three
bands nearly stick together at $\varepsilon _{p}{\rm \sim }\varepsilon _{d}$
with a cone-like behavior at the centre of the zone. As a result, the true
Wannier function of the anti-bonding band has very long range, but since $%
\varepsilon _{p}{\rm \sim }\varepsilon _{d}$ is 2--3 eV below the Fermi
level, the low-energy physics is hardly influenced by this. The second
reason why the orbital of interest cannot be a Wannier function, is that the
conduction band is crossed by, or has avoided crossings with other bands
(Fig.\thinspace \ref{CaCuO}). Since this occurs an eV below $\varepsilon
_{F},$ this, too, is irrelevant for the low-energy physics, which should
therefore be described using an orbital which yields correct wave functions
at and near $\varepsilon _{F}$ and has errors $\propto \left( \varepsilon
_{i}-\varepsilon _{F}\right) ^{N+1}$. The wider the energy range described
correctly by this orbital, {\it i.e.} the higher the $N,$ the longer its
spatial range.

We have found a {\em general} method, the NMTO method, by which for instance
this kind of orbital can be obtained\cite{NMTO}. What Fig.\thinspace \ref
{Figorb} shows is in fact a muffin-tin orbital (MTO) with $N$=1, obtained
from a density-functional (DF-LDA) NMTO calculation. This method has
recently enabled us to compute how the hopping integrals $t,$ $t^{\prime },$
and $t^{\prime \prime }$ are influenced by chemical and structural factors,
and it has proved successful for computing $t_{\parallel }$ and $t_{\perp }$
for the ladder cuprates without resort to the common, but dubious procedure
of fitting to guessed TB bands\cite{Ladder}.

In Fig.\thinspace \ref{CaCuO} we demonstrate that a {\em single} MTO of
sufficiently high $N$ {\em is} capable of describing the {\em entire}
conduction band, including its cone-like feature as well as smooth
interpolations across avoided crossings: The dotted band was obtained
variationally using an MTO with $N$=3, thus yielding band-errors of order $%
2(N+1)$=8. This figure also demonstrates that one may use a {\em discrete}
mesh of energies, $\epsilon _{0},...,\,\epsilon _{N},$ to construct the MTO,
which then has errors $\propto \left( \varepsilon _{i}-\epsilon _{0}\right)
...\left( \varepsilon _{i}-\epsilon _{N}\right) .$ This is analogous to
using Lagrange or Newton interpolation instead of Taylor expansion, and is
far more practical. The band obtained variationally has errors $\propto
\left( \varepsilon _{i}-\epsilon _{0}\right) ^{2}...\left( \varepsilon
_{i}-\epsilon _{N}\right) ^{2}.$

For some purposes, it is better to use a larger set of more localized
orbitals. For instance,{\it \ }in order to understand the microscopic
origins of $t,$ $t^{\prime },$ and $t^{\prime \prime }$, we used a set with
Cu$\,d_{x^{2}-y^{2}},$ O$\,p_{x}$, O$\,p_{y},$ and Cu$\,s,$\ obtained by
upfolding through a screening transformation\cite{NMTO,LMTO2,LMTO3}.

Materials with {\em many} bands and strong correlations are being studied
intensively. The first step of a quantitative description is a one-electron
mean-field theory requiring a basis, flexible enough to give individual
orbitals desired properties. For this, NMTOs are uniquely suited.

As an example of a minimal set spanning all states in a {\em wide} energy
range, let us consider the LDA valence and conduction bands for GaAs, 18 of
which fall in the range between -15\thinspace and +20\thinspace eV. With a Ga%
$\,sp^{3}d^{5}$ As$\,sp^{3}d^{5}f^{7}$ basis of merely 25 $N$=2 MTOs per
GaAs, and mesh points at -15, 0, and 10 eV, we obtained a variational band
structure, which only above +15 eV yielded errors as large as 0.1\thinspace
eV. Even for this 35\thinspace eV-range, which includes the Ga $3d$
semi-core band at -15 eV, {\em no} principal quantum numbers were needed. To
most practitioners, this is surprising result. NMTOs should be useful for
computing excited-state properties with the GW method\cite{GW}.

For ground-state properties, only the Ga $3d$ and the valence bands must be
described. Using the minimal Ga$\,sp^{3}d^{5}$ As$\,sp^{3}$ MTO set, we find
accuracies in the sum of the one-electron energies of 50 and 5 meV per GaAs
for respectively $N$=1 and $N$=2 \cite{NMTO}. This is highly satisfactory
and opens the way for accurate and efficient DF-calculations, for instance
for large systems using techniques where the computation increases merely
linearly with the size of the system. Hitherto, this has only been possible
with less accurate or geometry-restricted methods\cite{Odile,MRS}, such as
semi-empirical TB, screened LMTO-ASA\cite{LMTO2}, or screened
multiple-scattering theory\cite{Zeller}.

The LMTOs of the 1st- and 2nd-generations\cite{LMTO2} were expressed in
terms of partial waves, $\varphi _{Rl}\left( \epsilon _{0},r_{R}\right)
Y_{L}\left( \hat{r}_{R}\right) ,$ and their energy derivatives, $\dot{\varphi%
}_{Rl}\left( \epsilon _{0},r_{R}\right) Y_{L}\left( \hat{r}_{R}\right) ,$
truncated outside the atomic spheres $\left( r_{R}{\rm \equiv }\left| {\bf %
r-R}\right| \right) $. Everything else was neglected in the atomic-spheres
approximation (ASA), which then gave rise to a simple formalism and fast
computation. The 3rd-generation\cite{LMTO3} succeeds in making this
formalism valid for overlapping MT potentials, $V\left( {\bf r}\right) {\rm =%
}\sum_{R}v_{R}\left( r_{R}\right) ,$ to first order in the overlap of the $v$%
's, thus making the ASA superfluous. This is accomplished by attaching tails
of screened spherical waves with the proper energy to the partial waves. The
resulting set of {\em kinked} partial waves, evaluated on the energy mesh,
is what the NMTO set is expressed in terms of:
\begin{equation}
\chi _{R^{\prime }L^{\prime }}^{\left( N\right) }\left( {\bf r}\right)
=\sum\nolimits_{n=0}^{N}\sum\nolimits_{RL}\phi _{RL}\left( \epsilon _{n},%
{\bf r}\right) \,L_{nRL,R^{\prime }L^{\prime }}^{\left( N\right) }.
\label{e1}
\end{equation}
This may be considered as a polynomial approximation to the energy
dependence of the partial-wave set, in quantized form. In the following, we
derive the expressions for the Lagrange matrices, $L_{n}^{\left( N\right) },$
and the NMTO Hamiltonian and overlap matrices, starting out from the
conceptually simplest way of solving Schr\"{o}dinger's equation, namely by
matching of partial solutions. Our formalism should prove useful also in
other contexts.

We consider the case where the wave functions $\Psi _{i}\left( {\bf r}%
\right) $ are solutions of a Schr\"{o}dinger equation with a MT potential, $%
{\cal H}\Psi _{i}\left( {\bf r}\right) \equiv \left[ -\triangle +V\left( 
{\bf r}\right) \right] \Psi _{i}\left( {\bf r}\right) =\varepsilon _{i}\Psi
_{i}\left( {\bf r}\right) .$ For simplicity, we first assume that the MT
wells do not overlap and have ranges, $a_{R}.$ At the end, definitions will
be modified in such a way that the formalism holds also for overlapping
wells. The $a$'s will be hard-sphere radii which define the screening and,
hence, the shape of the orbitals.

{\it Kinked partial waves}\cite{LMTO3}. --Inside a MT sphere, the partial
solutions factorize into energy-dependent radial functions, $\varphi
_{Rl}\left( \varepsilon ,r_{R}\right) ,$ and angular functions. In the
interstitial, we use screened spherical waves, which are defined as those
solutions of the wave equation, $\left( \triangle +\varepsilon \right) \psi
_{RL}\left( \varepsilon ,{\bf r}\right) =0,{\bf \ }$which satisfy the
homogeneous boundary condition that the projection of $\psi _{RL}\left(
\varepsilon ,{\bf r}\right) $ onto $\delta \left( r_{R^{\prime
}}-a_{R^{\prime }}\right) Y_{L^{\prime }}\left( \hat{r}_{R^{\prime }}\right) 
$ be $\delta _{RR^{\prime }}\delta _{LL^{\prime }}.$ In fact, only those
solutions with $RL$ corresponding to the so-called {\em active} channels
will be used (in Fig.\thinspace \ref{Figorb}, the central Cu $d_{x^{2}-y^{2}}
$), and only the projections onto other {\em active} channels will vanish
(all non-central Cu $d_{x^{2}-y^{2}}$ projections). The projection of $\psi
_{RL}\left( \varepsilon ,{\bf r}\right) $ onto an {\em in}active channel
(all other than $d_{x^{2}-y^{2}}$ on any Cu-sphere) satisfies the boundary
condition that its radial logarithmic derivative equals that of the radial 
{\em Schr\"{o}dinger}-solution. The kinked partial wave, $\phi _{RL}\left(
\varepsilon ,{\bf r}\right) ,$ is now $\varphi _{Rl}\left( \varepsilon
,r_{R}\right) Y_{L}\left( \hat{r}_{R}\right) $ inside its own sphere and for
its own angular momentum, it is $\psi _{RL}\left( \varepsilon ,{\bf r}%
\right) $ in the interstitial region, and inside the sphere at ${\bf R}%
^{\prime },$ it vanishes for any other $\left( R^{\prime }L^{\prime }{\rm %
\neq }RL\right) $ active channel, but is proportional to $\varphi
_{R^{\prime }l^{\prime }}\left( \varepsilon ,r_{R^{\prime }}\right)
Y_{L^{\prime }}\left( \hat{r}_{R^{\prime }}\right) $ for an inactive
channel. As a result, with the normalization $\varphi _{Rl}\left(
\varepsilon ,a_{R}\right) {\rm \equiv }1,$ the kinked partial wave is a
continuous solution of Schr\"{o}dinger's equation with energy $\varepsilon .$
But since it has kinks at the spheres in the active channels, it is {\em not}
a wave function. 

The solid curve in the left-hand part of Fig.\thinspace \ref{KPW} shows the
Si $p_{x=y=z}$ kinked partial wave for $\varepsilon $ in the middle of the
valence band and for ${\bf r}$ along the [111]-line in the diamond structure
from the central Si atom, through the nearest Si neighbor, and half-way into
the back-bond void. The other curves will be explained when we come to
consider potential overlap. The kinks at the $a$-spheres (chosen smaller
than touching) are clearly seen. Since this kinked partial wave is designed
for use in a minimal $sp^{3}$-basis, only the Si $s$ and $p$ waves were
chosen as active. The inactive waves must therefore be provided by the tails
of the kinked partial waves centered at the neighbors, and this is the
reason for the strong Si $d$-character seen inside the nearest-neighbor
sphere. Had we been willing to keep Si $d$-orbitals in the basis, the Si $d$%
-channels would have been active so that only waves with $l$%
\mbox{$>$}%
2 would have remained inside the neighbor spheres, whereby the kinked
partial wave would have been more localized. Hence, the price for a smaller
kinked-partial wave basis, is longer spatial range and a stronger energy
dependence.

The element $K_{R^{\prime }L^{\prime },RL}\left( \varepsilon \right) $ of
the Hermitian {\em kink matrix\ }is defined as the kink of $\phi _{RL}\left(
\varepsilon ,{\bf r}\right) $ at the $a_{R^{\prime }}$-sphere, projected
onto $Y_{L^{\prime }}\left( \hat{r}_{R^{\prime }}\right) /a_{R^{\prime
}}^{2}.$ Hence, it specifies how the Hamiltonian operates on the set of
kinked partial waves: 
\begin{eqnarray}
\left( {\cal H}-\varepsilon \right) &&\phi _{RL}\left( \varepsilon ,{\bf r}%
\right) \equiv \left[ -\triangle +V\left( {\bf r}\right) -\varepsilon \right]
\phi _{RL}\left( \varepsilon ,{\bf r}\right) =  \label{e2} \\
&&-\sum\nolimits_{R^{\prime }L^{\prime }}\delta \left( r_{R^{\prime
}}-a_{R^{\prime }}\right) Y_{L^{\prime }}\left( \hat{r}_{R^{\prime }}\right)
K_{R^{\prime }L^{\prime },RL}\left( \varepsilon \right) .  \nonumber
\end{eqnarray}
Although an individual kinked partial wave is not a wave function, any {\em %
smooth} linear combination, $\sum_{RL}\phi _{RL}\left( \varepsilon ,{\bf r}%
\right) c_{RL,i}\,,$ is. Schr\"{o}dinger's equation may therefore be
formulated as the matching- or kink-cancellation condition: $%
\sum\nolimits_{RL}K_{R^{\prime }L^{\prime },RL}\left( \varepsilon
_{i}\right) c_{RL,i}=0$ for\ all\ $R^{\prime }L^{\prime },$ which is a set
of homogeneous linear equations, equivalent with the KKR equations\cite{MS}.
Here, the indices run only over active channels. Since the kink-matrix is
expensive to compute, it is not efficient to find a one-electron energy
from: $\det \left| K\left( \varepsilon _{i}\right) \right| {\rm =}0,$ and
then solve the linear equations for the corresponding $c_{RL,i}.$ Rather, we
construct a basis set, $\chi ^{\left( N\right) }\left( {\bf r}\right) ,$
with the property that it spans any wave function, $\Psi _{i}\left( {\bf r}%
\right) ,$ with an energy $\varepsilon _{i}$ in the neighborhood of $N$+1
chosen energies, $\epsilon _{0},...,\epsilon _{N},$ to within an error $%
\propto \left( \varepsilon _{i}-\epsilon _{0}\right) ...\left( \varepsilon
_{i}-\epsilon _{N}\right) $, and then solve the generalized eigenvalue
problem, 
\begin{equation}
\sum\nolimits_{RL}\left\langle \chi _{R^{\prime }L^{\prime }}^{\left(
N\right) }\left| {\cal H}-\varepsilon _{i}\right| \chi _{RL}^{\left(
N\right) }\right\rangle b_{RL,i}=0\quad {\rm for\ all}\ R^{\prime }L^{\prime
},  \label{e0}
\end{equation}
resulting from the Raleigh-Ritz variational principle.

{\it MTOs.} --\thinspace Since all wave functions with $\varepsilon _{i}{\rm %
=}\varepsilon $ may be expressed as: $\sum_{RL}\phi _{RL}\left( \varepsilon ,%
{\bf r}\right) c_{RL,i},$ the MTOs with $N$=0 are simply the kinked partial
waves at the chosen energy: $\chi _{RL}^{\left( 0\right) }\left( {\bf r}%
\right) {\rm =}\phi _{RL}\left( \epsilon _{0},{\bf r}\right) .$ The
Hamiltonian and overlap matrices are respectively $\left\langle \chi
^{\left( 0\right) }\left| {\cal H}-\epsilon _{0}\right| \chi ^{\left(
0\right) }\right\rangle =-K\left( \epsilon _{0}\right) $ and $\left\langle
\chi ^{\left( 0\right) }\mid \chi ^{\left( 0\right) }\right\rangle =\dot{K}%
\left( \epsilon _{0}\right) ,$ as may be found from Eq.\thinspace (\ref{e2})
and the normalization chosen. Here, $^{.}{\rm \equiv }\partial /\partial
\varepsilon $. In order to find the MTOs with $N$%
\mbox{$>$}%
0, we first define a {\em Green matrix:} $G\left( \varepsilon \right) \equiv
K\left( \varepsilon \right) ^{-1},$ and then, by an equation of the usual
type: $\left( {\cal H}-\varepsilon \right) \gamma _{RL}\left( \varepsilon ,%
{\bf r}\right) =-\delta \left( r_{R}-a_{R}\right) Y_{L}\left( \hat{r}%
_{R}\right) $, a Green function, $\gamma _{RL}\left( \varepsilon ,{\bf r}%
\right) ,$ which has one of its spatial variables confined to the $a$%
-spheres, {\it i.e.} ${\bf r}^{\prime }{\rm \rightarrow }RL.$ Considered a
function of ${\bf r,}$ this confined Green function is a solution with
energy $\varepsilon $ of the Schr\"{o}dinger equation, except at its own
sphere and for its own angular momentum, where it has a kink of size unity.
This kink becomes negligible when $\varepsilon $ is close to a one-electron
energy, because the Green function has a pole there. Eq.\thinspace (\ref{e2}%
) shows that $\gamma \left( \varepsilon ,{\bf r}\right) =\phi \left(
\varepsilon ,{\bf r}\right) G\left( \varepsilon \right) .$ (Here and in the
following, lower-case letters, such as $\gamma $ and $\phi ,$ denote
vectors, and upper-case letters, such as $K$ and $G$, denote matrices; $%
\varepsilon ,$ $\epsilon ,$ $RL,$ and $N$ are numbers, though). The confined
Green function is thus factorized into a Green matrix $G\left( \varepsilon
\right) $ which has the full energy dependence, and a vector of functions $%
\phi \left( \varepsilon ,{\bf r}\right) $ which has the full spatial
dependence and a weak energy dependence. (The energy windows we consider are
limited in size by the requirement that $\phi _{RL}\left( \varepsilon ,{\bf r%
}\right) $ and $\phi _{RL}\left( \varepsilon ^{\prime },{\bf r}\right) $
cannot be orthogonal). Finally, we want to factorize the ${\bf r}$ and $%
\varepsilon $-dependences completely and, hence, to approximate the confined
Green function by $\chi ^{\left( N\right) }\left( {\bf r}\right) G\left(
\varepsilon \right) :$ We note that subtracting from the Green function a
function which is analytical in energy, $\phi \left( \varepsilon ,{\bf r}%
\right) G\left( \varepsilon \right) -\omega ^{\left( N\right) }\left(
\varepsilon ,{\bf r}\right) \equiv \chi ^{\left( N\right) }\left(
\varepsilon ,{\bf r}\right) G\left( \varepsilon \right) ,$ produces an
equally good Green function in the sense that both yield the same Schr\"{o}%
dinger-equation solutions. If we can therefore determine the vector of
analytical functions, $\omega ^{\left( N\right) }\left( \varepsilon ,{\bf r}%
\right) ,$ in such a way that each $\chi _{RL}^{\left( N\right) }\left(
\varepsilon ,{\bf r}\right) $ takes the {\em same} value, $\chi
_{RL}^{\left( N\right) }\left( {\bf r}\right) ,$ at all mesh points, then $%
\chi _{RL}^{\left( N\right) }\left( \varepsilon ,{\bf r}\right) =\chi
_{RL}^{\left( N\right) }\left( {\bf r}\right) +O\left( \left( \varepsilon
-\epsilon _{0}\right) ..\left( \varepsilon -\epsilon _{N}\right) \right) .$
Hence, $\chi ^{\left( N\right) }\left( {\bf r}\right) $ is the set of NMTOs.
Now, since $\chi ^{\left( N\right) }\left( \epsilon _{0},{\bf r}\right) {\rm %
=...=}\chi ^{\left( N\right) }\left( \epsilon _{N},{\bf r}\right) $, the $N$%
th divided difference of $\chi ^{\left( N\right) }\left( \varepsilon ,{\bf r}%
\right) G\left( \varepsilon \right) $ equals $\chi ^{\left( N\right) }\left( 
{\bf r}\right) $ times the $N$th divided difference of $G\left( \varepsilon
\right) .$ Moreover, if we let $\omega ^{\left( N\right) }\left( \varepsilon
,{\bf r}\right) $ be a polynomial in energy of ($N$-1)st degree, its $N$th
divided difference on the mesh, $\Delta ^{N}\omega ^{\left( N\right) }\left( 
{\bf r}\right) /\Delta \left[ 0...N\right] ,$ will vanish. We have therefore
found the following solution: 
\begin{eqnarray}
\chi ^{\left( N\right) }\left( {\bf r}\right)  &=&\frac{\Delta ^{N}\phi
\left( {\bf r}\right) G}{\Delta \left[ 0...N\right] }\left[ \frac{\Delta
^{N}G}{\Delta \left[ 0...N\right] }\right] ^{-1}  \label{e3} \\[0.1cm]
&\equiv &\phi \left( \epsilon _{N},{\bf r}\right) +\frac{\Delta \phi \left( 
{\bf r}\right) }{\Delta \left[ N-1,N\right] }\left( E^{\left( N\right)
}-\epsilon _{N}\right) +..  \nonumber \\
&&..+\frac{\Delta ^{N}\phi \left( {\bf r}\right) }{\Delta \left[ 0...N\right]
}\left( E^{\left( 1\right) }-\epsilon _{1}\right) ..\left( E^{\left(
N\right) }-\epsilon _{N}\right) ,  \label{e4}
\end{eqnarray}
for the NMTO set. Since the kinks, $\left( {\cal H}-\varepsilon \right) \phi
\left( \varepsilon ,{\bf r}\right) G\left( \varepsilon \right) ,$ are
independent of $\varepsilon ,$ NMTOs with $N$%
\mbox{$>$}%
0 are smooth. By use of the well-known expression for a divided difference: 
\[
\frac{\Delta ^{N}\phi \left( {\bf r}\right) G}{\Delta \left[ 0...N\right] }%
=\sum_{n=0}^{N}\frac{\phi \left( \epsilon _{n},{\bf r}\right) G\left(
\epsilon _{n}\right) }{\prod_{m=0,\neq n}^{N}\left( \epsilon _{n}-\epsilon
_{m}\right) },
\]
we finally obtain the expressions for the Lagrange matrices in Eq.\thinspace
(\ref{e1}) and the energy matrices in Eq.\thinspace (\ref{e4}): $E^{\left(
M\right) }{\rm =}\left( \Delta ^{M}\varepsilon G/\Delta \left[ 0..M\right]
\right) \left( \Delta ^{M}G/\Delta \left[ 0..M\right] \right) ^{-1},$ in
terms of the values of the Green matrix on the energy mesh.

The NMTO set may thus be thought of as a 'quantized' Lagrange interpolation
of the kinked partial-wave set, where the weights are matrices rather than $N
$th-degree scalar polynomials in energy. Similarly, Eq.\thinspace (\ref{e4})
may be interpreted as a 'quantized' Newton interpolation with the energies
substituted by matrices. If the mesh is condensed$,$ Newton interpolation
becomes Taylor expansion: $\Delta ^{N}\phi /\Delta \left[ 0...N\right]
\rightarrow \left( 1/N!\right) d^{N}\phi /d\varepsilon ^{N}.$ The form (\ref
{e4}) expresses the NMTO as a kinked partial wave at the same site and with
the same angular momentum, plus a smoothing cloud of energy-derivative
functions centered at all sites and with all angular momenta. In the
right-hand part of Fig.\thinspace \ref{KPW}, the solid curve is the MTO with 
$N$=1, and the dashed curve is the MTO with $N$=0 shown also in the
left-hand part. Here again, longer spatial range is the price for spanning
the wave functions in a wider energy range. The increase of range and
smoothness with $N$ follows from the relation: $\left( {\cal H}-\epsilon
_{N}\right) \chi ^{\left( N\right) }\left( {\bf r}\right) =\chi ^{\left(
N-1\right) }\left( {\bf r}\right) \left( E^{\left( N\right) }-\epsilon
_{N}\right) ,$ which also shows that the $E$'s are transfer matrices between
MTO sets of different order. Linear transformations of the kinked partial
waves, $\hat{\phi}\left( \varepsilon ,{\bf r}\right) {\rm =}\phi \left(
\varepsilon ,{\bf r}\right) \hat{T}\left( \varepsilon \right) ,$ change the
NMTOs, but not the Hilbert space spanned by them\cite{NMTO}. This may be
used to generate nearly orthonormal representations where the $\hat{E}$'s
are Hamiltonians and where $\left\langle \hat{\chi}^{\left( M-1\right) }\mid 
\hat{\chi}^{\left( M\right) }\right\rangle \equiv 1$ for $1\leq M\leq N.$

The expressions for the Hamiltonian and overlap matrices needed in (\ref{e0}%
) may be worked out and given as\cite{NMTO}: 
\begin{eqnarray}
\frac{\Delta ^{N}G}{\Delta \left[ 0...N\right] } &&\left\langle \chi
^{\left( N\right) }\left| \varepsilon -{\cal H}\right| \chi ^{\left(
N\right) }\right\rangle \frac{\Delta ^{N}G}{\Delta \left[ 0...N\right] }=
\label{e5} \\
&&\frac{\Delta ^{2N}G}{\Delta \left[ \left[ 0..N-1\right] N\right] }+\left(
\varepsilon -\epsilon _{N}\right) \frac{\Delta ^{2N+1}G}{\Delta \left[ \left[
0...N\right] \right] }.  \nonumber
\end{eqnarray}
$\Delta ^{M+N+1}G/\Delta \left[ \left[ 0..M\right] N\right] $ is the ($M$+$N$%
+1)st derivative of that polynomial of degree $M$+$N$+1 which takes the
values $G\left( \epsilon _{0}\right) ,...,G\left( \epsilon _{N}\right) $ at
the $N$+1 mesh points and, at the first $M$+1 points, also the values $\dot{G%
}\left( \epsilon _{0}\right) ,..,\dot{G}\left( \epsilon _{M}\right) $ of the
energy-derivatives. The one-electron energies are 'ratios' of energy
derivatives of such 'Hermite interpolations' of $G\left( \varepsilon \right)
,$ which itself has poles inside the mesh.

Having seen that the formalism is expressed in terms of {\em one} matrix, 
{\it e.g. }$K\left( \varepsilon \right) {\rm =}\left\langle \chi ^{\left(
0\right) }\left| \varepsilon -{\cal H}\right| \chi ^{\left( 0\right)
}\right\rangle {\rm =}G\left( \varepsilon \right) ^{-1},$ let us indicate
how this is generated\cite{LMTO3,MSO}: The elements of the {\em bare} KKR
structure matrix\cite{MS}, $B_{R^{\prime }L^{\prime },RL}^{0}\left(
\varepsilon \right) \equiv \sum_{l"}4\pi i^{-l+l^{\prime }-l^{\prime \prime
}}C_{LL^{\prime }l^{\prime \prime }}\,\kappa n_{l^{\prime \prime }}\left(
\kappa \left| {\bf R-R}^{\prime }\right| \right) Y_{L^{\prime \prime
}}^{\ast }\left( \widehat{{\bf R-R}^{\prime }}\right) $ for $R{\rm \neq }%
R^{\prime },$ and ${\rm \equiv }0$ for $R{\rm =}R^{\prime },$ specify how
the spherical waves, $n_{l}\left( \kappa r_{R}\right) Y_{L}\left( \hat{r}%
_{R}\right) ,$ are expanded in regular spherical waves, $j_{l^{\prime
}}\left( \kappa r_{R^{\prime }}\right) Y_{L^{\prime }}\left( \hat{r}%
_{R^{\prime }}\right) .$ The corresponding expansions of the screened
spherical waves are now specified by a screened structure matrix, defined
via: $B^{\alpha }\left( \varepsilon \right) ^{-1}\equiv B^{0}\left(
\varepsilon \right) ^{-1}+\kappa ^{-1}\tan \alpha \left( \varepsilon \right)
,$ and obtained by matrix inversion of $B^{0}\left( \varepsilon \right)
+\kappa \cot \alpha \left( \varepsilon \right) .$ Here, $\kappa \cot \alpha
\left( \varepsilon \right) $ is a diagonal matrix with $\alpha _{RL}\left(
\varepsilon \right) $ being the hard-sphere phase shift, $\tan \alpha
_{Rl}\left( \varepsilon \right) {\rm \equiv }j_{l}\left( \kappa a_{R}\right)
/n_{l}\left( \kappa a_{R}\right) ,$ if the channel is active, and the true
phase shift, $\eta _{Rl}\left( \varepsilon \right) ,$ if the channel is
inactive. $B^{\alpha }\left( \varepsilon \right) $ has short spatial range
for energies well below the 'hard-sphere continuum,' as defined by the
division into active and inactive channels and the choice of $a$-radii for
the former. The kink matrix is finally: $K\left( \varepsilon \right) =-\left[
\kappa n\left( \kappa a\right) \right] ^{-1}\left[ B^{\alpha }\left(
\varepsilon \right) +\kappa \cot \eta ^{\alpha }\left( \varepsilon \right) %
\right] \left[ \kappa n\left( \kappa a\right) \right] ^{-1},$ where $\eta
^{\alpha }\left( \varepsilon \right) $ is the phase shift in the medium of
hard $a$-spheres:{\it \ }$\tan \eta ^{\alpha }\left( \varepsilon \right) 
{\rm \equiv }\tan \eta \left( \varepsilon \right) -\tan \alpha \left(
\varepsilon \right) .$ $B^{\alpha }\left( \varepsilon \right) $ contains the
essence of the hopping integrals, whose dependence on the local environment
enters through the screening.

When the potentials overlap, we need to redefine the kinked partial waves as
illustrated in Fig.\thinspace \ref{KPW}: $\phi _{RL}\left( \varepsilon ,{\bf %
r}\right) {\rm \equiv }\left[ \varphi _{Rl}\left( \varepsilon ,r_{R}\right) 
{\rm -}\varphi _{Rl}^{o}\left( \varepsilon ,r_{R}\right) \right] Y_{L}\left( 
\hat{r}_{R}\right) {\rm +}\psi _{RL}\left( \varepsilon ,{\bf r}\right) .$
Here, $\varphi \left( \varepsilon ,r\right) $ (dot-dashed) is the radial
solution for the central MT-well, which now extends to $s\,(>a).$ $\varphi
^{o}\left( \varepsilon ,r\right) $\ (dotted) is the phase-shifted wave
proceeding smoothly inwards from $s$ to the central $a$-sphere, where it is
matched with a kink to the screened spherical wave $\psi $ (dashed). It is
easily shown that, with this modification, the formalism holds to first
order in the potential-overlap\cite{NMTO,LMTO3,MSO}. In practice, this means
that radial overlaps of up to 30\% may be treated without changes, and that
overlaps as large as in Fig.\thinspace \ref{KPW}, may be treated by adding a
simple kinetic-energy correction\cite{NMTO,LMTO3,MSO,Catia}. This should
make the use of empty spheres superflous and open the way for efficient
DF-molecular-dynamics calculations. The $a$-radii now specify the screening,
with a default value which is 80\% of the atomic or ionic radius, and for
semi-core states, the core radius.

In conclusion, we have solved the long-standing problem of deriving useful,
minimal sets of short-ranged orbitals from scattering theory. Into a
calculation enters: (1) The phase shifts of the potential wells. (2) A
choice of which orbitals to include in the set, the so-called active
channels. (3) For these, a choice of screening radii, $a_{RL},$ to control
the orbital ranges. (4) An energy mesh on which the set will provide exact
solutions. These MTOs have significant advantages over those used in the
past.

\begin{figure}[b]
\caption{The Cu $d_{x^{2}-y^{2}}$-like LMTO, which describes the (LDA)
conduction band of HgBa$_{2}$CuO$_{4}$, plotted in the CuO$_{2}$ plane. Cu
and O sites are marked by respectively + and $\divideontimes .$}
\label{Figorb}
\end{figure}

\begin{figure}[th]
\caption{Band structure of CaCuO$_{2}$ with a 7$^{0}$-buckle, calculated in
the LDA with a single Bloch Cu $d_{x^{2}-y^{2}}$ CMTO (dotted) compared with
the full band structure (solid).}
\label{CaCuO}
\end{figure}

\begin{figure}[tbp]
\caption{Si $p_{x=y=z}$ kinked partial wave (KPW), its constituents $\protect%
\varphi ,\protect\varphi ^{o},$ and $\protect\psi $, and the LMTO. No empty
spheres were used. $s$ is the range of the central potential well.}
\label{KPW}
\end{figure}

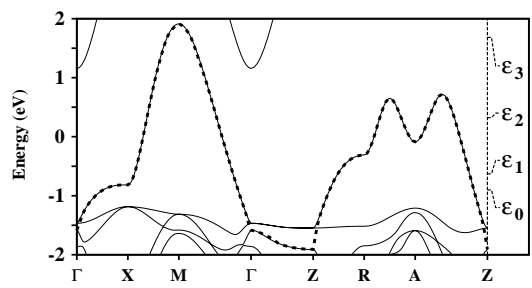
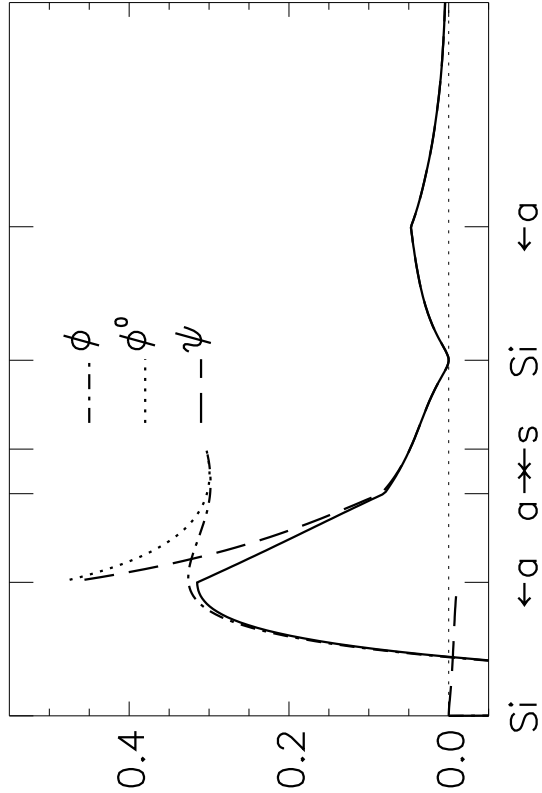
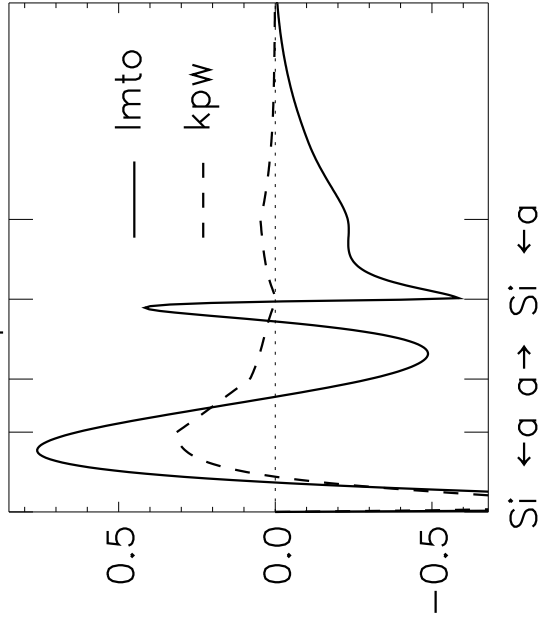
\end{document}